\documentclass[twocolumn]{jpsj2} 
\title{Optical conductivity of Ce-based filled skutterudites}

\author{
Tetsuya \textsc{Mutou}$^{1}$\thanks{E-mail address: tmutou@riko.shimane-u.ac.jp}  and
Tetsuro  \textsc{Saso}$^{2}$
}
\inst{
$^{1}$
Department of Material Science, Interdisciplinary Faculty of Science and Engineering,
Shimane University,
1060 Nishikawatsu-cho, Matsue-shi, Shimane 690-8504 \\
$^{2}$
Department of Physics, Faculty of Science, Saitama University,
Shimo-Ohkubo 255, Saitama 338-8570
}
\abst{
A simple tight-binding model is constructed for the description of the electronic structure of
some Ce-based filled skutterudite compounds showing an energy gap or pseudogap behavior. 
Assuming band-diagonal electron interactions on this tight-binding model,
the optical conductivity spectrum is calculated by applying the second-order self-consistent
perturbation theory to treat the electron correlation. The correlation effect is found to be of great importance
on the description of the temperature 
dependence of the optical conductivity. The rapid disappearance of  an optical gap
with increasing temperature is obtained as observed in the optical experiment for Ce-based
filled-skutterudite compounds.
}

\kword{filled skutterudite, optical conductivity, SCSOPT}

\begin{document}
\maketitle

\section{Introduction} 
In recent years, the filled skutterudite compounds with chemical formula RT$_{4}$X$_{12}$
(R = rare-earth elements, T = Fe, Ru and Os, and X = P, As, and Sb) have attracted much attention
because of several interesting physical properties including a new type of heavy-fermion superconductivity
observed in Pr-based compounds \cite{Bauer_02}.
In these filled skutterudites, some Ce-based compounds 
have received interest in relation to the Kondo insulators\cite{Riseborough00}. 
For CeRu$_{4}$Sb$_{12}$, for example,
it was reported that the (pseudo-)gap ($\sim 10 {\rm meV}$) opens 
in the spectra obtained by several measurements:
photoemission \cite{Kanai_01}, 
optical conductivity \cite{Dordevic_01} and inelastic neutron scattering\cite{Adroja_03,comment_1},
though the temperature dependence of resistivity shows metallic behavior \cite{Takeda_00}.
For CeOs$_{4}$Sb$_{12}$,  a semiconducting
behavior in the temperature dependence of resistivity was reported \cite{Bauer_01}, and a gap formation
($E_{\rm g} \sim 10 {\rm meV}$)
was observed in the optical measurement \cite{Matsunami_03}
at low temperatures.
In addition, a large peak at 70 meV,
which is called a mid-infrared (MIR) peak, was also observed.
With increasing temperature, the gap is gradually filled up. At high temperatures,
the peak structure loses its intensity and the gap disappears \cite{Matsunami_03}. 
The overall structure of the optical conductivity and the temperature dependence
of the spectrum are very similar to those of CeRu$_{4}$Sb$_{12}$ (MIR peak at 0.1eV)
in spite of the difference 
in the transport properties \cite{Dordevic_01}.
It should be noted that these overall structures are also found in a typical Kondo insulator YbB$_{12}$\cite{Okamura}
($E_{\rm g} \sim 20 {\rm meV}$ and MIR peak at 0.3 eV).

In fact, the overall band structure obtained by the band calculation for CeOs$_{4}$Sb$_{12}$ \cite{Harima_03}
is qualitatively similar to that for CeRu$_{4}$Sb$_{12}$ \cite{Harima_private},
though there is a discrepancy that the gap opens in CeRu$_{4}$Sb$_{12}$ which shows metallic temperature dependence in transport properties,
whereas it does not open in CeOs$_{4}$Sb$_{12}$ whose resistivity shows a semiconducting behavior. 
This discrepancy is not yet resolved and 
the difference in the transport properties might originate from unknown mechanism at low energies.  
They might be attributed to impurities or local strain.   In that case, refinement of the sample quality may resolve the discrepancy.
Nevertheless, one can expect that
similar structures of optical conductivity spectra for both Ce-based filled skutterudite compounds originate 
from similar band structures.

In this paper, we construct a simple tight-binding model which reproduces the overall structure of the 
band dispersion obtained by the band calculation for CeRu$_{4}$Sb$_{12}$,\cite{Harima_private}
and introduce the electron-electron interactions in the tight-binding model. We investigate the electron
correlation effect on the temperature dependence of optical conductivity spectra
and try to compare with the experiments.

\section{Model and Formulation}

Excluding T sites in RT$_{4}$X$_{12}$, 
the crystal structure of the filled skutterudites is regarded as
the body-centered cubic (bcc) structure consisting of X$_{12}$ icosahedrons with
a rare-earth ion R at a center. 
In CeOs$_{4}$Sb$_{12}$ or CeRu$_{4}$Sb$_{12}$, the f electron orbitals of Ce are expected to
hybridize with the f symmetry combination of the p orbitals of each Sb$_{12}$ cluster. 
Assuming an effective overlap integral between f wave functions on nearby R-sites
through Sb$_{12}$ clusters,
we consider a tight-binding model (with nearest-neighbor (ff$\sigma$)-bonds) \cite{Slater_54}
consisting of f symmetry wave functions on bcc lattice sites. 
By the spin-orbit interaction,  the 14-fold f states are split into the total angular momentum
$J=5/2$ and $J=7/2$ eigenstates.
Besides, under T$_{\rm h}$ symmetry, 
the $J=5/2$ states are split into $\Gamma_{5}$ and $\Gamma_{67}$ states,
and the $J=7/2$ states are split into two $\Gamma_{5}$'s and $\Gamma_{67}$.\cite{Takegahara_01}
Although overlap integrals (Slater-Koster parameters) for f electrons have been obtained in 
ref. \citen{Takegahara_79}
without the spin-orbit interaction, we need an expression in terms of $\Gamma_\alpha$ states for each $J$ value. This scheme was performed in ref.\citen{Saso_03} for YbB$_{12}$ and we follow the same procedure here. 
The resulting f bands, which will be shown in the next section, are rather similar to those above the Fermi level obtained by the LDA band calculation.\cite{Harima_private}

\begin{figure}[tb]
\begin{center}
\includegraphics[width=6cm]{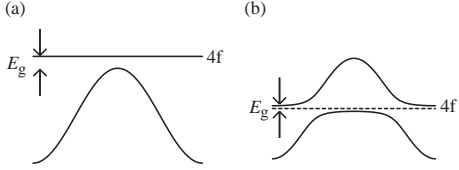}
\end{center}
\caption{
Schematic picture of the energy gap structure. (a) The present model. 
(b) The typical situation of the Kondo insulator.
}
\label{fig:energy_gap}
\end{figure}

The LDA bands below the Fermi level are rather complicated.
We found, however, that the top-most band (total width of 0.07eV) has a flat part of the width of about 0.02eV along
$\Gamma$-N-P-$\Gamma$ lines in the Brillouin zone, which is comparable to that of the f-bands above the Fermi level.
This may be partly due to the mixing with the f states above the gap since it contains a considerable f components.
Therefore, we express it by the simple tight-binding band of f symmetry states composed of Sb$_{12}$ p orbitals.
Note that the spin-orbit interaction can be neglected in this Sb$_{12}$ p band since
it is very weak in the p band.
%
Under T$_{\rm h}$ group,
the f symmetry states without the spin-orbit interaction are split into A$_{\rm u}$ and two T$_{\rm u}$'s.
In this paper, we assume that the Sb$_{12}$ p band with f A$_{\rm u}$ symmetry lies below the Fermi level and
the low-energy optical excitation originates from the transition from this to f bands
above the Fermi level for CeOs$_{4}$Sb$_{12}$ or CeRu$_{4}$Sb$_{12}$.
In analogy with semiconductors,
the former and the latter correspond to the valence and the conduction band, respectively.
It should be noted that the gap which we consider in the present model is not the hybridization gap
in contrast to the gap observed in other heavy fermion compounds
classified into the Kondo insulator / semiconductor,  e.g. YbB$_{12}$ \cite{Okamura}.
The difference of the (indirect) energy gap structure between the present model and the typical Kondo insulator 
is schematically shown in Fig.~\ref{fig:energy_gap}. In the present model, the 4f band (which is schematically shown
as an energy level in Fig.~\ref{fig:energy_gap}(a)) sits above the valence band,
and the gap opens between the 4f band and the valence band. 
On the other hand, in the typical Kondo insulator, the gap is constructed by the hybridization between
the 4f level and the conduction band (Fig.~\ref{fig:energy_gap}(b)). 
It should be noted that there should be some hybridization between the 4f and the valence bands also in the case (a).
In the present calculation, therefore, 
we assume the weak hybridization between the Sb$_{12}$ p band and f bands, 
hence our model lies between the cases (a) and (b).

We consider the effective tight-binding model mentioned above as a free system without
electron-electron interactions and assume effective Coulomb and exchange interactions as follows:\cite{Parmenter_73}
\begin{eqnarray}
\label{eqn:hamiltonian}
{\cal H}  &=& {\cal H}_{\rm band} + {\cal H}_{\rm int} , \\
\label{eqn:H_band}
{\cal H}_{\rm band} &=& 
\sum_{\gamma} \sum_{\mbox{\scriptsize\it\bf k}, \sigma}
E_{\mbox{\scriptsize\it\bf k}}^{\gamma}
c^{\dag}_{\gamma \mbox{\scriptsize\it\bf k} \sigma}c_{\gamma \mbox{\scriptsize\it\bf k} \sigma}, \\
{\cal H}_{\rm int} &=& U\sum_{\gamma}\sum_{i}
n_{\gamma i \uparrow}n_{\gamma i \downarrow} 
+ U_{2}\sum_{\gamma < \gamma'}\sum_{i, \sigma}
n_{\gamma i \sigma}n_{\gamma' i \bar{\sigma}} \nonumber \\
& & + U_{3}\sum_{\gamma < \gamma'}\sum_{i, \sigma}
n_{\gamma i \sigma}n_{\gamma' i \sigma} \nonumber \\
& & -J\sum_{\gamma < \gamma'}\sum_{i, \sigma}
c^{\dag}_{\gamma i \sigma}c_{\gamma i \bar{\sigma}}
c^{\dag}_{\gamma' i \bar{\sigma}}c_{\gamma' i \sigma},
\label{eqn:H_int} 
\end{eqnarray}
where
$E_{\mbox{\scriptsize\it\bf k}}^{\gamma}$ denotes the diagonalized tight-binding band energy 
for the band $\gamma$. 
We denote the annihilation (creation) operator for the band $\gamma$
in the site representation as $c_{\gamma i \sigma}$ ($c^{\dag}_{\gamma i \sigma}$). 
Here the index $\sigma$ denotes a pair of the time-reversal states which we call spin hereafter.
We assume the the {\it para} state for the spin and omit the index $\sigma$.
In the Hamiltonian, we have introduced the above form of electron interactions expressed by the parameters $U$, $U_{2}$, $U_{3}$ and $J$.
They correspond to the intra-band Coulomb interaction, the inter-band anti-parallel spin Coulomb
interaction, the inter-band parallel spin Coulomb interaction, and the inter-band exchange interaction,
respectively.
In the interaction parts ${\cal H}_{\rm int}$ of the Hamiltonian, we assume band-diagonal
electron interactions.  Usually the interactions are introduced between the  localized 
electron orbitals before the diagonalization of $E_{\mbox{\scriptsize\it\bf k}}^{\gamma}$, 
but other choices of basis functions are also possible in general.
They are equivalent with each other, but a difference may arise when one neglects some of the matrix elements.
The present form is suitable for the calculation of the many-body effect, as will be seen below. 
Besides, in the present paper, the interaction parameters are chosen to fit the experiments, 
so that a different choice of the interaction form might not change our result. 

In the Hamiltonian (\ref{eqn:H_int}), the Coulomb interactions are introduced in all bands
since there is the hybridization between the Sb$_{12}$ p band and f bands,
although
the strong electron-electron interactions should exist only between the f electrons in real systems.
However, the equal Coulomb interaction may act more strongly on the region of high density of states, so that the f-components will be more affected than the others.
For simplicity and in order to reduce the number of parameters,
the following relations are further assumed; $U_{2}=U-J$ and
$U_{3}=U-2J$ \cite{comment_2}. 
In this case, ${\cal H}_{\rm int}$ can be expressed as
\begin{eqnarray}
{\cal H}_{\rm int} &=&
\frac{U}{2}\sum_{\gamma\gamma'\sigma\sigma'}\sum_{i}
c^{\dag}_{\gamma' i \sigma'} c^{\dag}_{\gamma i \sigma}
c_{\gamma i \sigma} c_{\gamma' i \sigma'} \nonumber \\
&-& \frac{J}{2}\sum_{\gamma\gamma'\sigma\sigma'}\sum_{i, \sigma}
c^{\dag}_{\gamma' i \sigma'}c^{\dag}_{\gamma i \bar{\sigma}}
c_{\gamma i \bar{\sigma'}}c_{\gamma' i \sigma},
\end{eqnarray}
which has a rotational symmetry in both the orbital and the spin states.

\begin{figure}[tb]
\begin{center}
\includegraphics[width=6cm]{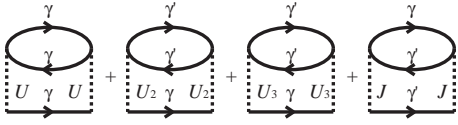}
\end{center}
\caption{
Second-order perturbation terms of the self energy $\tilde{\Sigma}^{\gamma}(\varepsilon + {\rm i}\delta)$ for
the $\gamma$ component. The solid line denotes
the full local Green's function. Dashed lines indicate electron-electron
interactions: $U$, $U_{2}$, $U_{3}$ or $J$.
}
\label{fig:selfenergy}
\end{figure}

In the present model, it is noted that 
the electron correlation effect beyond the Hartree-Fock approximation does not appear in the ground state at
$T=0$.
In order to take account of the correlation effect, 
we apply the self-consistent second-order perturbation theory (SCSOPT) to the present model with
the local approximation for the self-energy part of the Green's function. It is expected that
the SCSOPT is sufficient to investigate the correlation effect in the low-energy and low temperature region.
Using the density of states (DOS) $D^{\gamma}(\nu)$ for each band $\gamma$,
the local Green's function is expressed as follows;
\begin{eqnarray}
\label{eqn:Green}
G^{\gamma}(\varepsilon + {\rm i} \delta) &=&
\int {\rm d}\nu \frac{\displaystyle D^{\gamma}(\nu)}
{\displaystyle \varepsilon + {\rm i} \delta-\nu-\Sigma^{\gamma}(\varepsilon + {\rm i} \delta)}, \\
D^{\gamma}(\nu) &=&  \frac{\displaystyle 1}{\displaystyle N}
\sum_{\mbox{\scriptsize\it\bf k}}\delta(\nu-E_{\mbox{\scriptsize\it\bf k}}^{\gamma}),
\end{eqnarray}
where $N$ is the number of sites and $\delta\rightarrow 0^+$.
The self energy $\Sigma^{\gamma}(\varepsilon + {\rm i} \delta)$ consists of
the constant Hartree term
and the second-order perturbation term $\tilde{\Sigma}^{\gamma}(\varepsilon + {\rm i} \delta)$. 
The former is neglected here since it can be regarded as already included in the band calculation.
The latter is calculated 
from the full local Green's function as shown in Fig.~\ref{fig:selfenergy}. 
In these diagrams, the solid line denotes the 
full local Green's function $G^{\gamma}(\varepsilon + {\rm i} \delta)$ 
which is determined self-consistently together with the self-energy. 

The number density of electrons is denoted by $n^{\gamma}$:
\begin{eqnarray}
\label{eqn:number}
n^{\gamma} &=& 2 \int {\rm d}\varepsilon \rho^{\gamma}(\varepsilon) f(\varepsilon), \\
\rho^{\gamma}(\varepsilon) &=& -\frac{\displaystyle 1}{\displaystyle \pi}
{\rm Im}G^{\gamma}(\varepsilon + {\rm i} \delta),
\end{eqnarray}
where $f(\varepsilon)$ is the Fermi distribution function 
: $f(\varepsilon) = 1/({\rm e}^{\beta (\varepsilon-\mu)} +1)$
and $\beta = (k_{\rm B}T)^{-1}$.
The chemical potential $\mu$ should be determined so that $n_{\rm total}( = \sum_{\gamma}n^{\gamma})$ is equal to
the given total density of electrons.
In the practical calculation,  we set the chemical potential at $T=0$ in the middle point between 
the top of the valence band and the bottom of the conduction band, and we use the fixed value of
the chemical potential at any temperature for simplicity.

According to the linear response theory, the optical conductivity is obtained from the current-current
correlation function (eq.(\ref{eqn:direct}) in Appendix). 
In deriving the formula,
we neglect $\mbox{\boldmath$k$}$ and ${\gamma}$ dependence 
of the velocity matrix element of the current operator.
The transition of electrons in the optical absorption is the direct one conserving the momentum.
We, however, found that the formula eq.(\ref{eqn:direct}) can not reproduce the experiments at all because a sharp threshold behavior appears at the direct gap edge (see Fig.~\ref{fig:comparison} in Appendix).
We therefore assume that the momentum conservation is violated in the real system
because of imperfections and phonon-assisted transitions. 
By these simplifications, 
the current-current correlation function is reduced to the joint-DOS-type form \cite{Urasaki_99}.
In the present paper,
we use the following expression for the optical conductivity by omitting constant factors:
\begin{equation}
\label{eqn:optical}
\sigma(\omega) \equiv \sum_{\gamma, \gamma'}
\int {\rm d}\varepsilon
\rho^{\gamma}(\varepsilon)\rho^{\gamma'}(\varepsilon+\omega)
\frac{\displaystyle f(\varepsilon)-f(\varepsilon+\omega)}{\displaystyle \omega}.
\end{equation}
There is a problem that the joint-DOS-type form does not give the correct spectrum for 
$\omega \rightarrow 0$ at $T > 0$. In the present study, however, we use the joint-DOS-type
form of the optical conductivity since we treat the system with the gap and investigate the overall
structure of the spectrum.

\section{Results}

\begin{figure}[tb]
\begin{center}
\includegraphics[width=6cm]{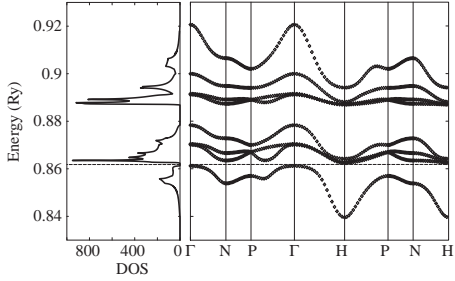}
\end{center}
\caption{
The total DOS (left) and the band structure (right) for the non-interacting system. The Fermi level
is indicated by the dashed line.
}
\label{fig:FREE}
\end{figure}

Figure~\ref{fig:FREE} shows the band dispersion and the total DOS we use in the case without
interactions. The bands above the Fermi level (indicated by the dashed line) are the f bands
constructed by the tight-binding model with the two-center integral $({\rm ff}\sigma)=0.005$ Ry.
In order to reproduce the band dispersions obtained by the LDA band calculation,
we choose $E_{\Gamma_{5}}=0.870$ Ry and $E_{\Gamma_{67}}=0.867$ Ry
for $J=5/2$.
The band structure just above the Fermi level corresponds to $J=5/2$
and that in the high-energy region corresponds to $J=7/2$.
$J=7/2$ states have matrix elements for the crystalline electric field (CEF) term $(O_6^2-O_6^6)$ in the T$_{\rm h}$ group, but we neglected them for simplicity since the $J=7/2$ states lie at high energies.  $E_{\Gamma_{5}^{(1)}}$, $E_{\Gamma_{5}^{(2)}}$ and $E_{\Gamma_{67}}$ are chosen as
$0.894$ Ry, $0.902$ Ry and $0.889$ Ry, respectively.
These values are chosen independently from the CEF parameters since there might be contributions due to hydridization from implicit surrounding orbitals (not included in the present model explicitly).
For the valence band corresponding to Sb$_{12}$ p band with the A$_{\rm u}$ symmetry, 
the value of $({\rm ff}\sigma)$ is set equal to $0.002$ Ry and the energy level to $0.857$ Ry.
The hybridization between the Sb$_{12}$ p band and f bands is introduced by (ff$\sigma)'$, which is set equal to $0.003$ Ry.
There is an indirect gap between $\Gamma$ and H points in accord with the LDA band.  
The position of the direct gap differs from the $\Gamma$ point in the LDA band,
but as will be seen later, we consider the indirect gap as most important.

Concerning the valence band, there is a rather flat part between N and P points.
This is an effect of the hybridization (ff$\sigma)'$ between the valence band and the 4f band above $E_F$ and yield a peak in the DOS.
The total valence band width assumed in the present calculation is rather narrow than that estimated from the band calculation \cite{Harima_03}. 
As mentioned in the previous section, we have regarded the present band with the 
narrow width and the above-mentioned peak at 0.855 Ry as corresponding to the sharp peak structure at the top of the valence band obtained from the band calculation, and the contributions from the bands below it are neglected.
It may be allowed since the main contribution to the low-energy excitation spectrum of the optical conductivity originates from the  bands near the gap.

\begin{figure}[tb]
\begin{center}
\includegraphics[width=6cm]{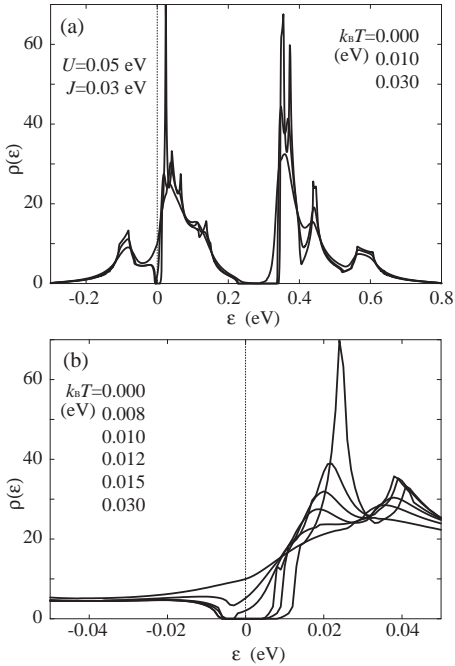}
\end{center}
\caption{
(a) Temperature dependence of the total DOS $\rho(\varepsilon)$.
(b) Expanded figure of the total DOS near the gap.
}
\label{fig:DOS_Tdep}
\end{figure}

In the present calculation, we set $U=0.05$eV and $J=0.03$eV. We choose these values
to reproduce the temperature dependence of the optical conductivity in the experiment. 
Though the DOS does not have any temperature dependence within the Hartree approximation,
the strong temperature dependence shows up if the correlation effect is included. 
Figure~\ref{fig:DOS_Tdep}(a)
shows the temperature dependence of the total DOS $\rho(\varepsilon)$ (it is noted 
that the chemical potential is set to the origin in the figures of the total DOS).
The sharp peaks lose their intensity and their structures
become smooth as the temperature increases. 
Note that the temperature-dependent spectrum of the DOS cannot be obtained without taking account
of the correlation effect.
The temperature dependence near the gap is shown in Fig.~\ref{fig:DOS_Tdep}(b).
One can see that the gap is filled up with increasing temperature and it turns into the pseudogap at the
high temperatures comparable to the gap size. 
The peak structures near the gap lose their intensity and they shift to the center of the gap.
In the present model we use, the gap at $T=0$ is equal to about 0.02 eV.
Thus it is recognized that the gap is filled in at much lower temperature than that corresponding to
the gap size itself. At the high temperatures corresponding to the gap size, the gap structure disappears.

\begin{figure}[tb]
\begin{center}
\includegraphics[width=6cm]{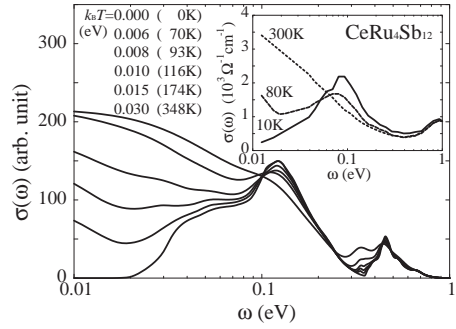}
\end{center}
\caption{
Temperature dependence of the optical conductivity calculated by using the SCSOPT. 
Inset shows the experimental data 
of CeRu$_{4}$Sb$_{12}$ \cite{Dordevic_01}. Peaks due to phonons observed in the gap are omitted.
Note that the logarithmic scale is used for $\omega$ axis.
}
\label{fig:DYN_Tdep}
\end{figure}

The temperature dependence of the optical conductivity is shown in 
Fig.~\ref{fig:DYN_Tdep}. 
The spectrum has a large peak near $\omega \simeq 0.1$ eV and 
the (indirect) gap which reflects the DOS gap at 0.02eV.
With increasing temperature, the intraband contribution (the Drude part)  appears and the 
spectrum has finite intensity in the gap region. The large peak loses its intensity with increasing
temperature. At the highest temperature we calculated ($k_{\rm B}T=0.030$ eV, which is higher than
the room temperature), the gap and the peak structures disappear completely.
In the optical experiment for CeRu$_{4}$Sb$_{12}$ \cite{Dordevic_01}, the gap is filled up with 
increasing temperature and it disappears completely at 300K as shown in the inset of Fig.~\ref{fig:DYN_Tdep}.
In the experimental data, one can see also the decrease of the intensity of the MIR peak with increasing 
temperature.

\begin{figure}[tb]
\begin{center}
\includegraphics[width=6cm]{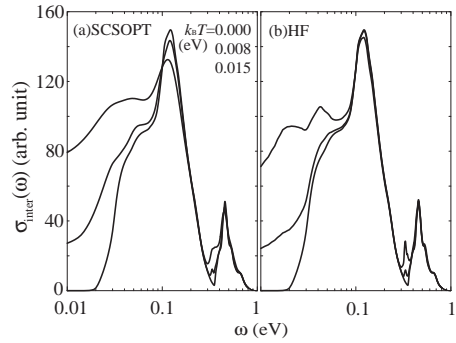}
\end{center}
\caption{
Comparison of the interband contributions of optical conductivity spectra
calculated by using (a) the SCSOPT and (b) the Hartree-Fock (HF) approximation. 
}
\label{fig:DYN_comparison}
\end{figure}

In order to see the correlation effect in the temperature dependence of the optical conductivity more clearly,
we show a comparison of spectra calculated by using the SCSOPT and the Hartree-Fock (HF) approximation
in Fig.~\ref{fig:DYN_comparison}. 
Note that the figure shows the interband contribution $\sigma_{\rm inter}(\omega)$ which consists of
the part $\gamma \neq \gamma'$ of the summation in eq.(\ref{eqn:optical}).
Obviously the rapid decrease of the intensity corresponding to the
MIR peak can not be obtained within the HF approximation. 
Since the temperature dependence
of the optical conductivity calculated by using the HF approximation originates from only 
the Fermi distribution function,  
the large peak of the spectrum does not show any remarkable change with increasing temperature.

\section{Summary and Discussion}

We have calculated the density of states and the optical conductivity for
the simple tight-binding model, including local Coulomb and exchange interactions by SCSOPT;
thereby the correlation effect on the temperature dependence of optical conductivity spectra
has been investigated.
The intensity in the gap region grows rapidly and the large peak above the gap shifts
to the lower energy with increasing temperature and the gap
disappears at high temperatures; 
the gap is filled at the lower temperature than that corresponding to the gap size itself.  
In the optical measurement for  CeOs$_{4}$Sb$_{12}$ \cite{Matsunami_03} and
CeRu$_{4}$Sb$_{12}$ \cite{Dordevic_01}, the gap structure disappears at a lower temperature 
than the gap size. 
The rapid disappearing of the gap and the shift of
the peak structure cannot be obtained without the correlation effect. Thus it is concluded
that the correlation effect is important to explain the temperature dependence of
the optical conductivity spectra obtained by the experiments for the Ce-based filled 
skutterudites : CeOs$_{4}$Sb$_{12}$ and CeRu$_{4}$Sb$_{12}$. In the photoemission
experiment for CeRu$_{4}$Sb$_{12}$ \cite{Kanai_01}, the temperature dependence of 
the photoemission spectrum was investigated. In their result,
the pseudo-gap is filled with increasing temperature. The temperature-dependent DOS spectrum 
cannot be explained without the correlation effect as mentioned repeatedly. 

In the present study, it has been shown that the large peak corresponding to the MIR peak
originates from the optical transition to the $J=5/2$ states from the valence band.
Thus, there should be the another peak corresponding to the transition to the $J=7/2$ states
at higher energy regions. In fact, in the present result, we can obviously see a peak structure
corresponding to the transition to the $J=7/2$ state at about $0.5$ eV in Fig.~\ref{fig:DYN_Tdep}. 
When we consider that the large peak (MIR peak) near $0.1$ eV observed in experimental data
for CeOs$_{4}$Sb$_{12}$ and CeRu$_{4}$Sb$_{12}$ originates in the optical transition
from the valence band of X$_{12}$ clusters to the conduction
band consisting of dispersive f $J=5/2$ bands, it can be expected that there is another peak structure
corresponding to the transition to f $J=7/2$ bands.
In the experimental result for CeRu$_{4}$Sb$_{12}$
\cite{Dordevic_01}, there is a peak around $1$ eV (inset of Fig.~\ref{fig:DYN_Tdep}). 
For CeOs$_{4}$Sb$_{12}$ \cite{Matsunami_03},
there seems to be also a peak structure near $1$ eV. According to the band calculation \cite{Harima_03},
the magnitude of the spin-orbit splitting can be estimated as $0.02 \sim 0.03$ Ry, i.e. a few hundreds meV.
Thus the peak position near $1$eV in the experimental spectrum is too high and one 
can not simply conclude that it originates from the transition to $J=7/2$ band. 
Although this problem cannot be readily explained by the present study, 
it should be reconciled whether there is a possibility that
the transition to $J=7/2$ corresponds to the peak structure near  $1$ eV
in the optical conductivity spectrum for CeOs$_{4}$Sb$_{12}$ or CeRu$_{4}$Sb$_{12}$ both from experimental and theoretical sides.
In order to discuss the detail structure of the spectrum, however, we need a further study
with a more elaborated band-structure model.

Finally, we comment that a similar theoretical scheme adopted in the present paper was recently applied to the optical spectra of the typical Kondo insulator YbB$_{12}$\cite{Saso04} with a lot of common features, indicating an underlying shared structures in both materials.

\section*{Acknowledgments}

We would like to thank Professor H. Harima for providing us the band calculation data, and Professor H. Okamura and Dr. M. Matsunami for informative discussions
on recent experimental results.
A main part of the numerical calculations was performed on the 
super-computer VPP700E of RIKEN (The Institute of Physical and Chemical Research).
One of the authors (T.S.) is indebted to the Grant-in-Aid for Scientific Research on
 (B)(1), ``Development of General Band Calculation Program by the Dynamical Mean-Field Thory'', No. 14340108 and 
Priority Areas, ``Evolution of New Quantum Phenomena Realized in the Filled Skutterudite Structure'', No. 16037204
from the Ministry of Education, Science and Culture.

\appendix

\section{Comment on the joint-DOS-type optical conductivity}

\begin{figure}[tb]
\begin{center}
\includegraphics[width=6cm]{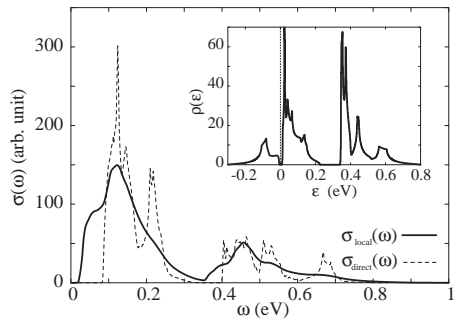}
\end{center}
\caption{
Comparison of the optical conductivity spectra obtained from eq.(\ref{eqn:optical}) 
$\sigma_{\rm local}(\omega)$ and that from eq.(\ref{eqn:direct}) $\sigma_{\rm direct}(\omega)$
in the case without interactions (see text). Inset shows the total DOS without interactions in the eV energy unit.
}
\label{fig:comparison}
\end{figure}

In this Appendix, we comment on the comparison of the optical conductivity spectrum
obtained from the local DOS (eq.(\ref{eqn:optical})) and that obtained from the direct optical
transition.  When the momentum conservation
is assumed in derivation of the optical conductivity, we obtain the following expression for the optical
conductivity with the direct optical transition (assuming velocity matrix elements to be constant and neglecting the vertex correction again):
\begin{eqnarray}
\label{eqn:direct}
\sigma_{\rm direct}(\omega) &\equiv& \frac{\displaystyle 1}{\displaystyle N}
\sum_{\mbox{\scriptsize\it\bf k}}
\sum_{\gamma, \gamma'}
\int {\rm d}\varepsilon \rho^{\gamma}_{\mbox{\scriptsize\it\bf k}}(\varepsilon)
\rho^{\gamma'}_{\mbox{\scriptsize\it\bf k}}(\varepsilon+\omega) \nonumber \\
&\times& \frac{\displaystyle f(\varepsilon)-f(\varepsilon+\omega)}{\displaystyle \omega}, 
\end{eqnarray}
\begin{equation}
\label{eqn:k-Green}
\rho^{\gamma}_{\mbox{\scriptsize\it\bf k}}(\varepsilon)
= -\frac{\displaystyle 1}{\displaystyle \pi}{\rm Im}
\frac{\displaystyle 1}{\displaystyle \varepsilon + {\rm i}\delta  - 
E^{\gamma}_{\mbox{\scriptsize\it\bf k}} - \Sigma^{\gamma}(\varepsilon + {\rm i}\delta)},
\end{equation}
where we omit the constant factor in defining $\sigma_{\mbox{direct}}(\omega)$.
In contrast to $\sigma_{\rm direct}(\omega)$,
we denote here the joint-DOS-type optical conductivity expressed by eq.(\ref{eqn:optical})
as $\sigma_{\rm local}(\omega)$.
The comparison of $\sigma_{\rm local}(\omega)$ and $\sigma_{\rm direct}(\omega)$ for the non-interacting
case is shown in Fig.~\ref{fig:comparison}. 
The gap appearing in $\sigma_{\rm local}(\omega)$
reflects the indirect gap and it is less than the direct gap reflected in $\sigma_{\rm direct}(\omega)$, but the overall structure including positions of large peaks is consistent with each other. 

For $T > 0$, the gap of DOS is smeared out by the many-body effect. In this case, the Drude part for $\gamma = \gamma'$
becomes as 
$\sigma_{\rm Drude}^{\gamma}(\omega \rightarrow 0) = \frac{2 \tau}{\pi}
\rho^{\gamma}(0)$ from (\ref{eqn:direct}) 
if $\rho^{\gamma}_{\mbox{\scriptsize\it\bf k}}(\varepsilon) = -\frac{1}{\pi}
{\rm Im}
\frac{1}{\varepsilon-E^{\gamma}_{\mbox{\scriptsize\it\bf k}}+{\rm i}/(2\tau)}$, 
although it becomes $(\rho^{\gamma}(0))^{2}$ from (\ref{eqn:optical}).
Therefore, it should be noted that the joint-DOS-type form of the optical conductivity
does not give the correct spectrum for $\omega \rightarrow 0$ at $T > 0$ and $\rho^\gamma(0)>0$.
Despite this deficiency, $\sigma_{\rm local}(\omega)$ can explain the gross feature of the observed spectra at finite frequencies in the realistic situations.


\begin{thebibliography}{99}
\bibitem{comment_1}
Strictly speaking, the opening of the gap in the spectrum of optical and magnetic excitation
is not observed in experiments for CeRu$_{4}$Sb$_{12}$, 
though a disappearance of the spectrum intensity in a lower energy region was expressed as the "gap formation"
in literatures.
\bibitem{comment_2}
In fact, there is no reason to assume these relations in the present model.
In a special case, however, these relations between electron interactions
are essential to satisfy the rotational invariance in real and spin spaces \cite{Urasaki_99,Parmenter_73}.
\end{thebibliography}
\end{document}